\documentclass[12pt,a4paper]{amsart}

\usepackage{amsaddr}
\usepackage{amssymb}
\usepackage[final]{showkeys}
\usepackage{microtype}
\usepackage{color}
\usepackage{graphicx}
\usepackage[hmargin=3cm,vmargin={3.5cm,4cm}]{geometry}

\numberwithin{equation}{section}

\allowdisplaybreaks

\begin{document}

    \title[G-fractional diffusion with boundaries]{On G-fractional diffusion models in bounded domains}

\author{L. Angelani$^{1,2}$ and R. Garra$^{3}$}
\address{$^1$ ISC-CNR, Institute for Complex Systems, P.le A. Moro 2, 00185 Rome, Italy}
\address{$^2$ Dipartimento di Fisica, Sapienza Universit\`a di Roma, P.le A. Moro 2, 00185 Rome, Italy}
\address{$^3$ Institute of Marine Sciences, National Research Council (CNR),
Via del Fosso del Cavaliere, I-00133 Rome, Italy}


    \date{\today}

    \begin{abstract}
     
    In the recent literature, the g-subdiffusion equation involving Caputo fractional derivatives with respect to another function has been studied in relation to anomalous diffusions with a continuous transition between different subdiffusive regimes. 
    In this paper we study the problem of g-fractional diffusion in a bounded domain with absorbing boundaries. We find the explicit solution for the initial-boundary value problem and we study the first passage time distribution and the mean first-passage time (MFPT). 
    The main outcome is the proof that with a particular choice of the function $g$ it is possible to obtain a finite MFPT, differently from the anomalous diffusion described by a fractional heat equation involving the classical Caputo derivative. 
    \end{abstract}

    \maketitle
	
	\section{Introduction}

	Anomalous diffusions described by g-fractional differential equations have been investigated in a series of recent papers (see e.g. \cite{kosto} and \cite{kosto1}) in relation to subdiffusive models with a continuous transition between two different regimes.
    We recall that g-fractional diffusions are fractional heat-type equations involving time-fractional derivatives with respect to another function $g$. These generalized fractional derivatives can be obtained by means of a deterministic change of variable and include as special cases the Hadamard and the Erd\'elyi-Kober derivatives (see \cite{kilbas}).
     However, some of the consequences of this change of variable in the definition of the integro-differential operator appearing in the governing equations are not trivial. First of all, g-fractional derivatives are useful to describe different anomalous diffusions such as ultra-slow processes \cite{ale}.
     Moreover, these equations provide interesting fractional-type generalizations of classical PDEs with variable coefficients. 
     Heuristically g-fractional derivatives are useful in order to take into account in a single integro-differential operator the memory effect and the time-dependence of the diffusivity (see for example \cite{garra} for the application to the Dodson-type diffusion).
    
    \bigskip
    
    In this paper we investigate g-fractional diffusions in bounded 
    and semi-bounded
    domains with absorbing boundaries. Different papers have been devoted to fractional equations involving Caputo derivatives in bounded domains, with explicit representation of the solution and the probabilistic meaning, we refer for example to \cite{agrawal,beghin,fa,phys2000}. 
    Here we provide an explicit representation of the solution for the initial-boundary value problem and discuss the role of the $g$-function for the computation of the mean first-passage time (MFPT). In particular, we show that there is a choice of $g$ functions such that the MFPT turns out to be finite, unlike classical fractional diffusion based on Caputo derivatives. 
    We also report some graphs showing the trend of the 
    numerically evaluated first-passage time distributions
%    We also provide numerical simulations of the First passage-time distributions 
    for some interesting choices of the function $g$, highlighting the main differences in their asymptotic behaviors.
    We finally show how the known solution of the g-fractional diffusion in unbounded space is obtained as a limit of our expressions.
    The present analysis and results are relevant in order to better understand the role played by the function $g$ in fractional diffusive models.

    \section{G-fractional diffusion in bounded domains with absorbing boundaries}
    
    In a series of recent papers \cite{kosto, kosto1}, the authors have discussed the utility for anomalous diffusion models of the so-called g-fractional derivatives (also named in the literature fractional derivatives w.r.t. another function \cite{kilbas} or $\psi$-fractional derivative \cite{sousa}).
   In other recent papers, some particular form of the g-fractional diffusive equations have been considered in relation to interesting models. For example in \cite{ale} for ultra-slow diffusions and in \cite{garra} for the generalized Dodson equation.

    Here we consider a fractional diffusion in a bounded domain with absorbing boundaries. 
    We recall that the g-fractional derivative of order $\alpha \in (0,1)$ is defined as
    \begin{equation}
    \left(\frac{^C\partial^\alpha_gu}{\partial t^\alpha} \right)(x,t) = \frac{1}{\Gamma(1-\alpha)} \int_0^t (g(t)-g(\tau))^{-\alpha}\frac{\partial u}{\partial \tau} d\tau,
    \end{equation} 
where $g$ is a deterministic function such that $g(0) = 0 $ and $g'(t)>0$ for $t>0$, where we denote by $g' = dg/dt$ the first order time derivative. This fractional operator can be obtained by means of a change of variable from the classical 
Caputo derivative and includes interesting cases such as the Hadamard derivative (for $g = \ln(t+1)$) and the Erd\'elyi-Kober derivative (for $g = t^\beta$). Obviously by taking
$g = t$ we recover the definition of the Caputo derivative. \\
We also observe that 
the solution of the fractional equation 
\begin{equation}
\left(\frac{^C d^\alpha_g f}{d t^\alpha} \right)(t) = -\lambda f(t),
 \end{equation}
under the initial condition $f(0)= 1$, is given by
\begin{equation}
f(t) = E_{\alpha}\left(-\lambda g^\alpha\right)	 = \sum_{k=0}^\infty \frac{(-\lambda g^\alpha)^k}{\Gamma(\alpha k+1)},
\end{equation}
where $E_\alpha(\cdot)$ denotes the one-parameter Mittag-Leffler function \cite{ml}. 
Therefore, in this case, an eigenfunction of the fractional derivative is given by the Mittag-Leffler function composed with the function $g(t)$.

\subsection{Fractional diffusion with two absorbing boundaries}

Let us consider the g-fractional diffusion equation 
\begin{equation}
\frac{^C\partial^\alpha_gu}{\partial t^\alpha} = D \frac{\partial^2 u}{\partial x^2},
\end{equation}
in the bounded domain $x \in [-a,b]$, $a, b>0$, under the following initial and boundary conditions
\begin{equation}
u(x,0) = \delta (x-x_0), \quad u(-a, t) = 0, \quad u(b,t)=0,
\end{equation}
corresponding to a particle performing an anomalous diffusion 
(with generalized diffusion constant $D$)
in a bounded domain with absorbing boundaries. \\
It is possible to find a solution by means of the separation of variable method, i.e.
$$u(x,t) = X(x)T(t).$$
We have to solve the equations
\begin{align}
&\label{1}\frac{^C d^\alpha_g T(t)}{d t^\alpha}  = -\lambda^2 D T(t),\\	
& \label{2}\frac{d^2 X(x)}{dx^2} = -\lambda^2 X(x),
\end{align}
where $\lambda^2$ is the separation constant.

The solution of the equation \eqref{2} under these boundary conditions is given by 
\begin{equation}
	X(x) = A_n \sin[\lambda_n (x+a)], 
\end{equation}
with 
$$ \lambda_n = \frac{n\pi}{a+b},$$
corresponding to the eigenvalue problem with the given conditions.
By using the fact that the Mittag-Leffler function 
$E_\alpha(-\lambda_n^2 D g^\alpha)$ provides the solution of the time-fractional equation \eqref{1} and by combination of the space and time component of the solution we have
\begin{equation}
		u(x,t) = \sum_{n=1}^\infty A_n \sin[\lambda_n (x+a)]E_\alpha(-\lambda_n^2 D g^\alpha).
\end{equation}
The coefficient $A_n$ can be found imposing the initial condition and we finally find that the explicit form of the solution of the initial-boundary value problem is 
\begin{equation}
	u(x,t) = \sum_{n=1}^\infty \frac{2\sin[\lambda_n(x_0+a)]\sin[\lambda_n(x+a)]}{a+b}E_\alpha(-\lambda_n^2 D g(t)^\alpha).
\label{ufin}
\end{equation}
We observe that, for $\alpha = 1 $ we have
\begin{equation}
	u(x,t) = \sum_{n=1}^\infty \frac{2\sin[\lambda_n(x_0+a)]\sin[\lambda_n(x+a)]}{a+b}\exp(-\lambda_n^2 D g),
\end{equation}
that is the solution of a diffusion equation with variable diffusivity
\begin{equation}
\frac{\partial u}{\partial t} = Dg'(t)\frac{\partial^2 u}{\partial x^2},
\end{equation}
including for example the diffusive equation governing the fractional Brownian motion.

\begin{figure}[t!]
\includegraphics[width=.48\linewidth] {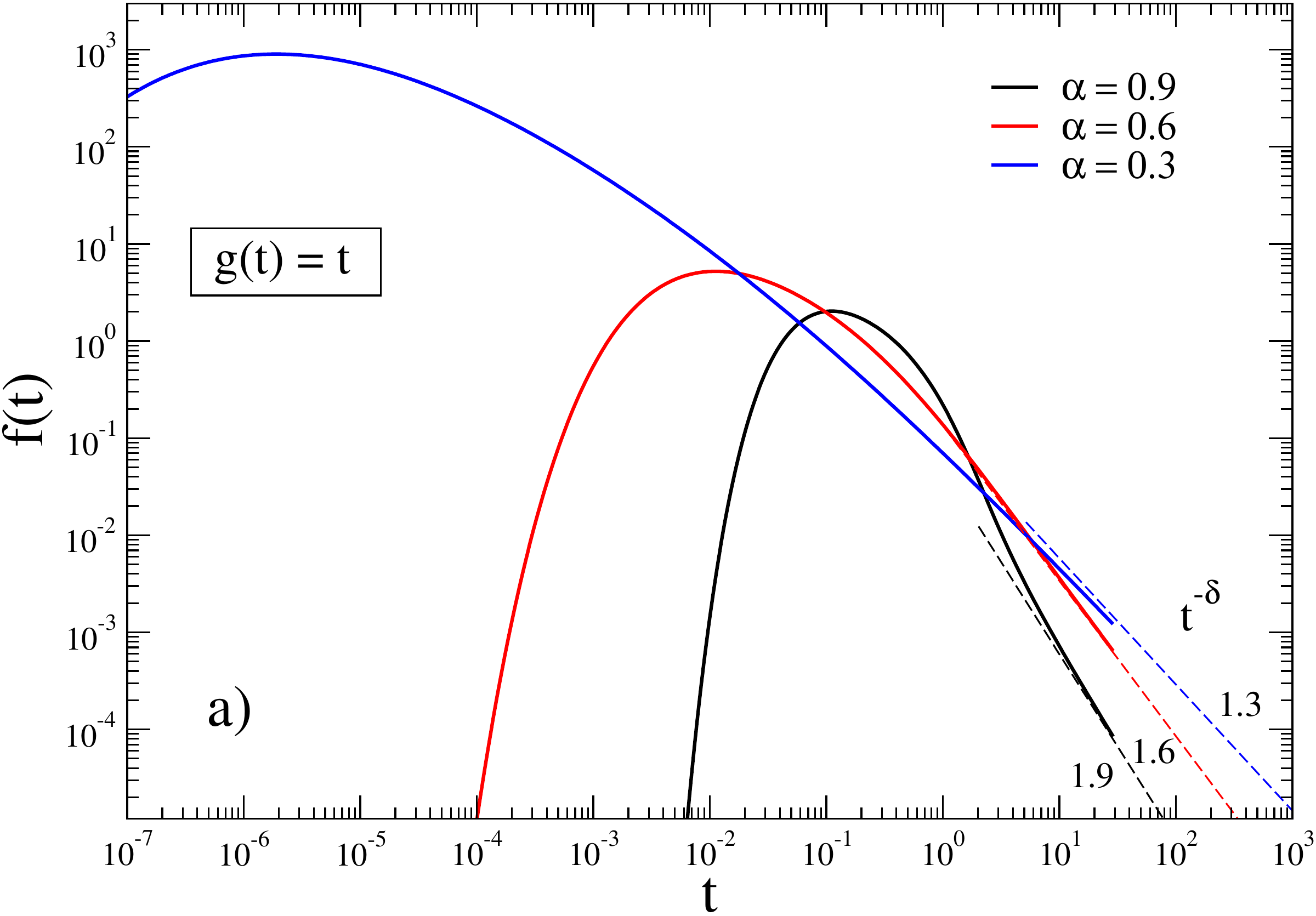}
\includegraphics[width=.48\linewidth] {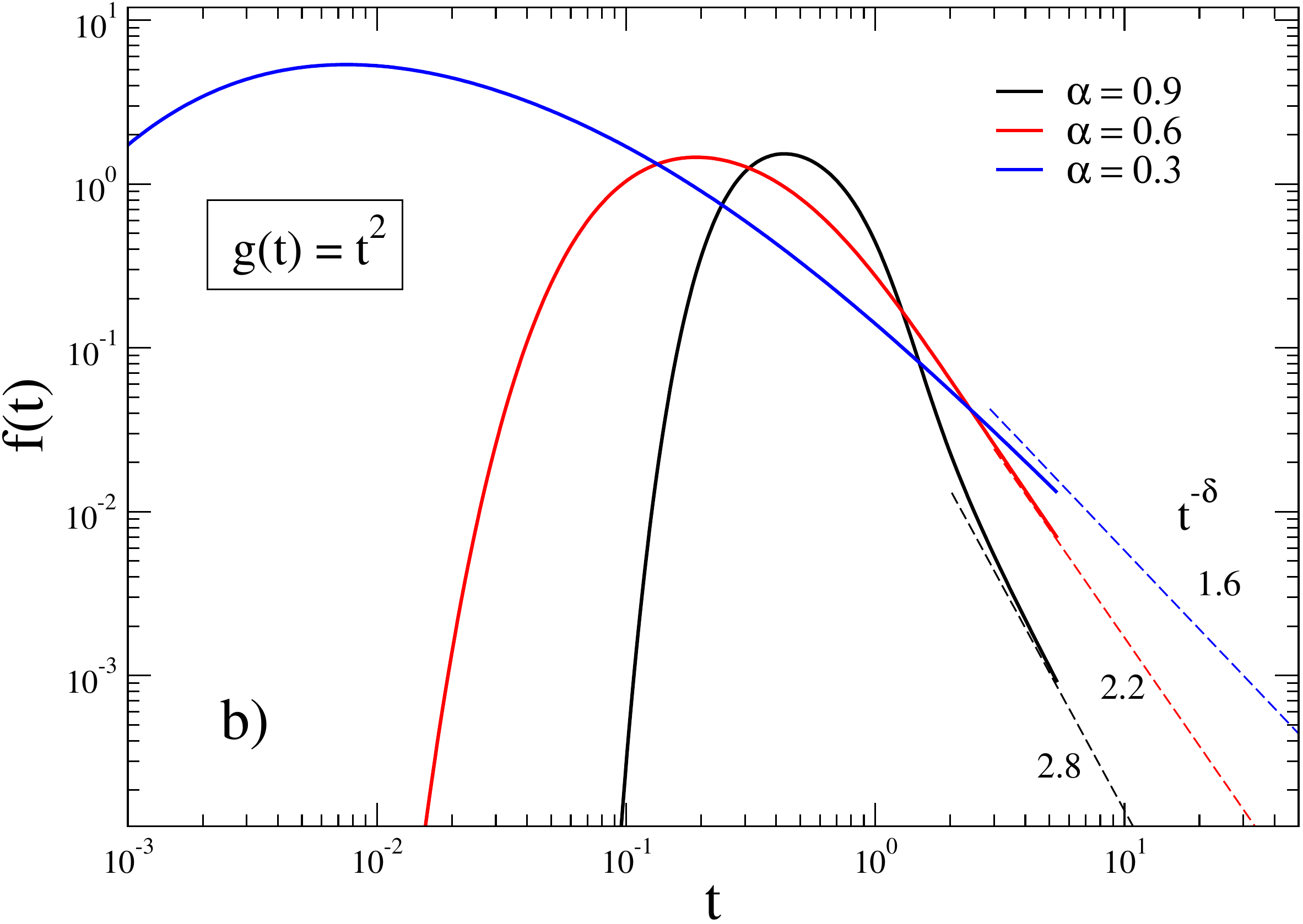}
\includegraphics[width=.48\linewidth] {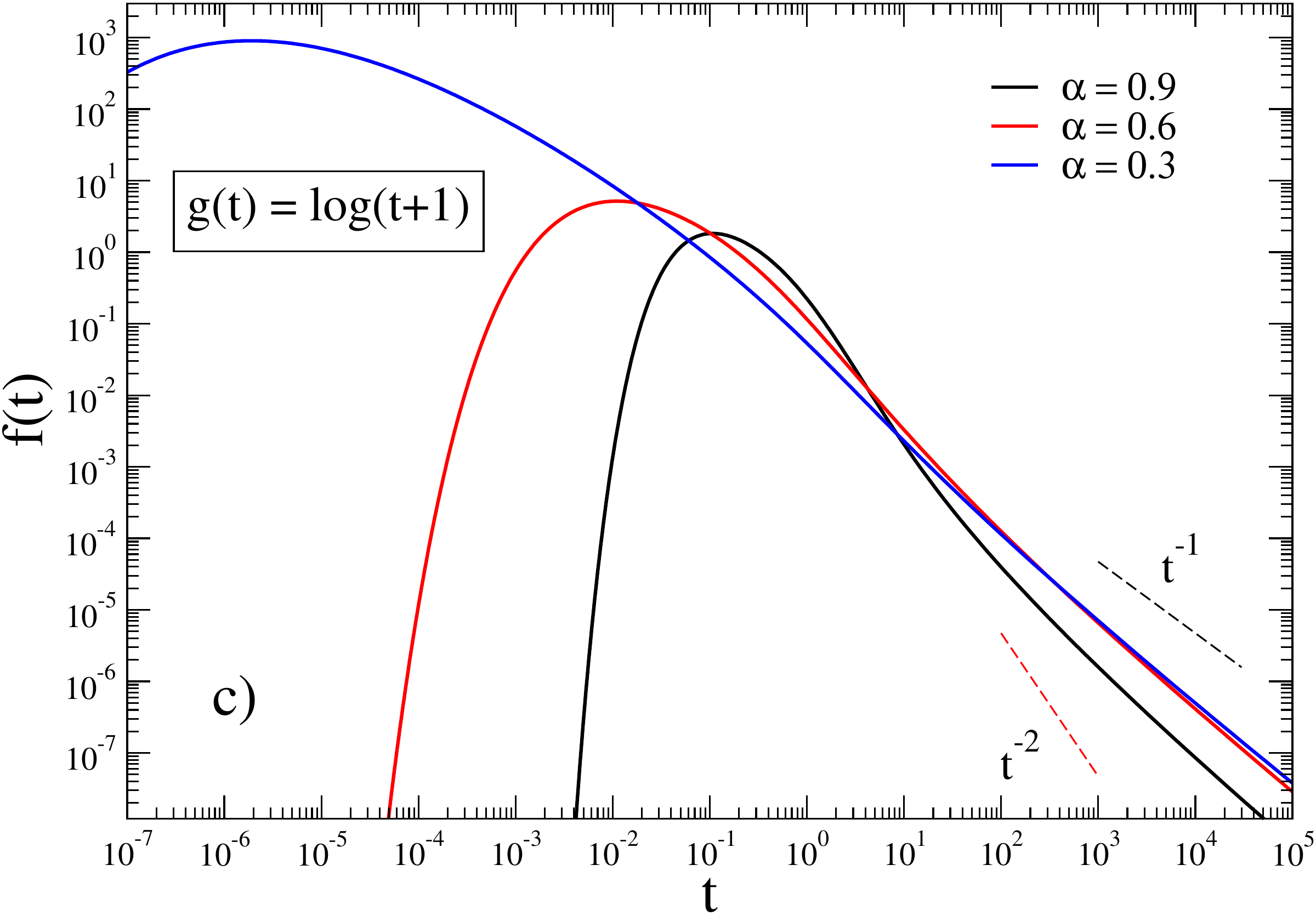}
\includegraphics[width=.48\linewidth] {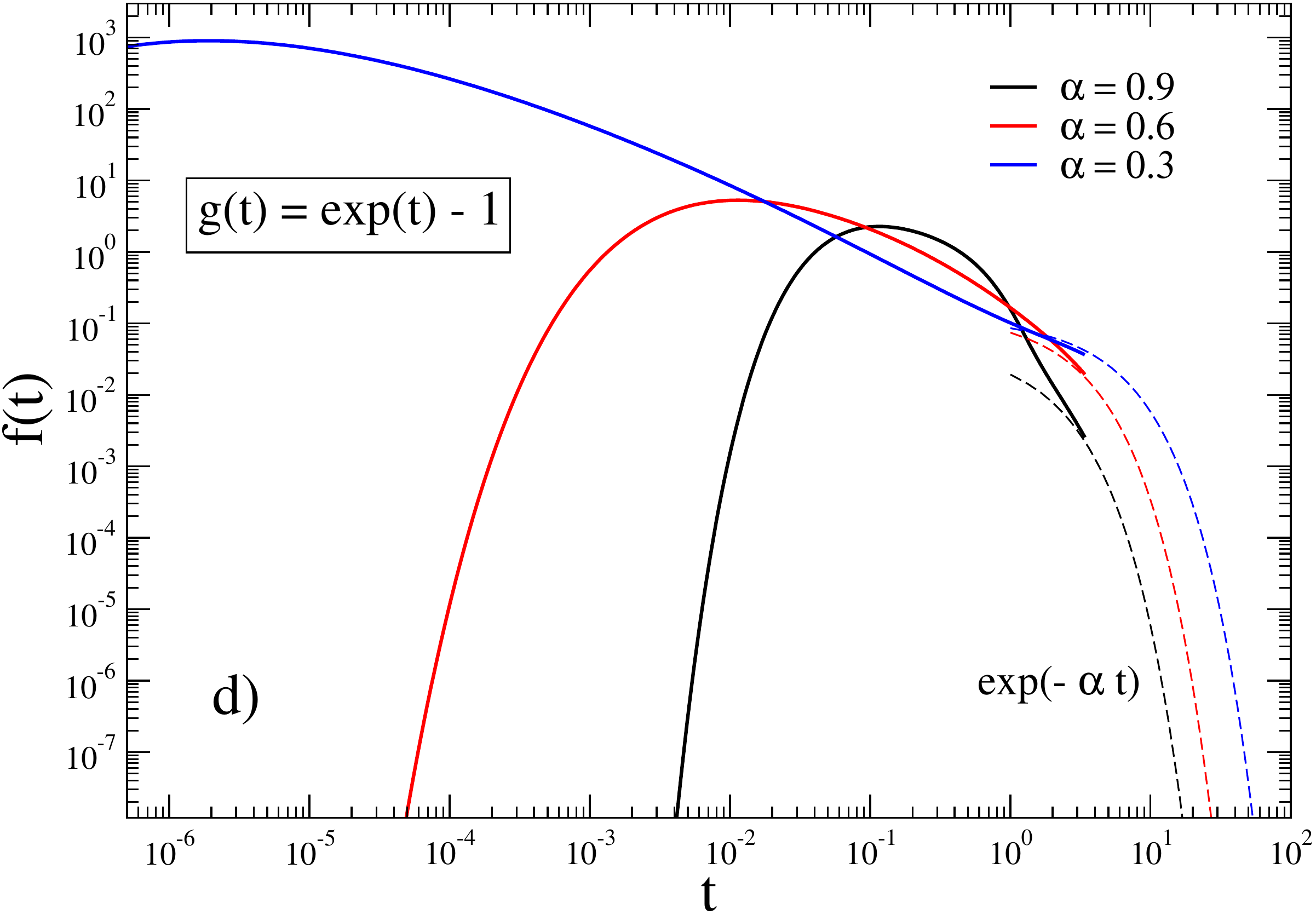}
\caption{First passage-time distributions for different choices of the function $g$.\\
a) The case of Caputo fractional derivative, $g(t)=t$.
Dashed lines are asymptotic expressions (\ref{f3asyn}),
$f = A \ t^{-\delta}$ with $\delta=1+\alpha$. 
The MFPT is infinite for any value of $\alpha$.\\
b) The case of Erd\'elyi-Kober derivative, $g(t)=t^\beta$ with $\beta=2$. The asymptotic behavior exponent is $\delta=1+\alpha \beta$. Only $\alpha>1/2$ corresponds to finite MFPT.\\
c) The case of Hadamard fractional derivative, $g(t)=\ln(t+1)$. The asymptotic behavior is $t^{-1} \ln{(t)}^{-1-\alpha}$, between $t^{-1}$ and $t^{-2}$ (guide for eyes), implying divergent MFPT.\\
d) The case of exponential derivative $g(t)=\exp(t)-1$. Asymptotic expressions are given by $A \exp(-\alpha t)$ and the MFPT is always finite.\\
In all panels we report FPTD (full lines) calculated numerically inverting the Laplace transform (\ref{phicap}) and using (\ref{f1}), for three different values of the fractional derivative, $\alpha=0.9,0.6,0.3$. Dashed lines are asymptotic expressions (\ref{fasyn}). We set $x_0=0$, $a=b=1$ and $D=1$, so the prefactor in the asymptotic expressions is $A=-1/(2\Gamma(-\alpha))$.
}
\label{fig1}
\end{figure}

We now study the first passage time distribution (FPTD) as a function of $g(t)$ in order to underline the main difference w.r.t. to the time-fractional model considered in \cite{ranga}.
The FPTD $f(t)$ can be calculated as follows
\begin{equation}
f(t) = -\frac{d}{dt}\int_{-a}^b dx \ u(x,t).
\end{equation}
We recall that
\begin{equation}
\int_{-a}^b \sin \left(\frac{n\pi(x+a)}{a+b}\right)dx = \frac{2(a+b)}{n\pi} \quad \mbox{if n is odd},
\end{equation}
and null otherwise. Therefore, by substitution, we have that
\begin{equation}
f(t) = -\frac{d}{dt}\frac{4}{\pi}\sum_{n=0}^\infty \frac{1}{2n+1}
\sin\left(\frac{(2n+1)\pi(a+x_0)}{a+b}\right)E_{\alpha}\left(-\frac{(2n+1)^2\pi^2}{(a+b)^2}Dg(t)^\alpha\right).
\end{equation}
We now observe that 
\begin{align}
\nonumber \frac{d}{dt}E_{\alpha}\left(-\frac{(2n+1)^2\pi^2}{(a+b)^2}D g(t)^\alpha\right) &= \sum_{k=0}^\infty\frac{ \bigg(-\frac{(2n+1)^2D\pi^2}{(a+b)^2}\bigg)^k}{\Gamma(\alpha k+1)}\frac{d}{dt}g(t)^{\alpha k}\\
& = \sum_{k=1}^\infty\frac{ \bigg(-\frac{(2n+1)^2D\pi^2}{(a+b)^2}\bigg)^k g'(t)g(t)^{\alpha k-1}}{\Gamma(\alpha k)}\nonumber\\
& = \sum_{k=0}^\infty\frac{ \bigg(-\frac{(2n+1)^2D\pi^2}{(a+b)^2}\bigg)^{k+1} g'(t)g(t)^{\alpha k+\alpha-1}}{\Gamma(\alpha k+\alpha)}\nonumber\\
& = -\frac{(2n+1)^2\pi^2 Dg'(t)g(t)^{\alpha-1}}{(a+b)^2}E_{\alpha, \alpha}\bigg(-\frac{(2n+1)^2\pi^2}{(a+b)^2}Dg(t)^{\alpha}\bigg)\nonumber,
\end{align}
where we have used $\Gamma(z+1)=z\Gamma(z)$
and introduced the two-parameters Mittag-Leffler function (see e.g. \cite{ml})
%we denoted by $g' = dg/dt$ and 
$$E_{\alpha, \beta}(x)= \sum_{k=0}^\infty \frac{x^k}{\Gamma(\alpha k+\beta)}.$$
We finally have that the FPTD is given by 
\begin{equation}
	f(t) = \frac{4\pi D g'(t){g(t)^{\alpha-1}}}{(a+b)^2}\sum_{n=0}^{\infty}
	(2n+1)\sin\bigg[\frac{(2n+1)\pi(a+x_0)}{a+b}\bigg]E_{\alpha, \alpha}\bigg(-\frac{(2n+1)^2\pi^2}{(a+b)^2}Dg(t)^{\alpha}\bigg).
\label{FPTD}
\end{equation}
For $g(t) = t$ we recover the result obtained in \cite{ranga}. 
The previous expression is very general, being valid for generic functions $g$. 
In the following we will discuss some interesting case studies, such as the classical Caputo derivative $g(t)=t$, the Erdélyi-Kober derivative $g(t)=t^\beta$, the Hadamard derivative $g(t)=\ln(t+1)$ and the exponential derivative $g(t)=\exp(t)-1$, 
also reporting some typical behaviors in Fig.\ref{fig1}.

%It is interesting for example the ultra-slow case $g(t) = \ln(1+t)$.

We now study the mean first-passage time (MFPT), analyzing the conditions under which it has finite values. 
%%%  NEW
The MFPT is defined as the first moment of the distribution $f(t)$
\begin{equation}
	\tau = \int_0^\infty dt \ t \ f(t).
	\label{MFPT}
\end{equation}
In order to determine the conditions for the existence of a finite MFPT we have to investigate the asymptotic behavior of the FPTD (\ref{FPTD}). By using the asymptotic expansion of the Mittag-Leffler function for $|z| \to \infty$ and $\Re(z)<0$
(see \cite{ml}, p. 75)
\begin{equation}
E_{\alpha, \alpha}(z)= - \frac{z^{-2}}{\Gamma(-\alpha)} 
+ O(|z|^{-3}),
\end{equation}
and considering that long times means large values of $g$
(due to the constraint $g'(t)>0$), we have that the asymptotic behavior of (\ref{FPTD}) reads
\begin{equation}
f(t) \sim A\ g'(t) \ g(t)^{-(1+\alpha)},
\hspace{1cm} t \to \infty,
\label{fasyn}
\end{equation}
where the time-independent prefactor $A$ is
\begin{equation}
	A = - \frac{4 (a+b)^2}{\pi^3 D \Gamma(-\alpha)}\sum_{n=0}^{\infty}
	(2n+1)^{-3}\sin\bigg[\frac{(2n+1)\pi(a+x_0)}{a+b} \bigg]. \label{FPTDlongt}
\end{equation}
We observe that, in the symmetric case, $a=b$ and $x_0=0$,
the quantity $A$ takes the simple form $A=-a^2/(2D\Gamma(-\alpha))$.
The condition for the existence of a finite MFPT reduces then to a condition on the asymptotic behavior of the function $g$,
which, from (\ref{MFPT}) and (\ref{fasyn}), must satisfy 
\begin{equation}
\lim_{t \to \infty} t^2 \ \cdot  g'(t) \ \cdot g(t)^{-(1+\alpha)} = 0.
\label{condex}
\end{equation}
Before discussing the above condition for different choices of $g$, let us first note that it would also have been possible to obtain the asymptotic behavior ({\ref{fasyn}})  by using known results about anomalous diffusion processes with classical Caputo derivative \cite{JPA,Git}. 
Indeed, the FPTD (\ref{FPTD}) can be expressed as
\begin{equation}
	f(t) = g'(t) \ \varphi(g(t)),
	\label{f1}
\end{equation}
where $\varphi(t)$ is the first-passage time distribution of the problem with Caputo derivative, i.e. $g(t)=t$.
The asymptotic behavior of $\varphi(t)$ is obtained from 
the known expression of its Laplace transform 
(see Eq.(36) in \cite{JPA} or Eq.(7) in \cite{Git})
\begin{equation}
\tilde{\varphi}(s) = \int_0^\infty dt \ e^{-st}\ \varphi(t) =
\frac{\cosh{\frac{c(b-a-2x_0)}{2}}}{\cosh{\frac{c(b+a)}{2}}},
\label{phicap}
\end{equation}
where we denoted with $\tilde{\varphi}(s)$ the Laplace transform of $\varphi(t)$,  $s$ is the Laplace variable and  $c^2=s^\alpha /D$. For small $s$ we have that 
$\tilde{\varphi}(s) \sim 1 - a\ s^\alpha$.
By using the Tauberian theorem for the survival probability 
$\tilde{\mathbb{P}}(s) = [1-\tilde{\varphi}(s)]/s$, we have that 
$\tilde{\mathbb{P}}(s)\sim s^{\alpha-1}$ for small $s$ and then 
$\mathbb{P}(t)\sim t^{-\alpha}$ for large $t$ \cite{ksbook}.
By differentiation we finally deduce the asymptotic behavior 
$\varphi(t) \sim t^{-(1+\alpha)}$, that, inserted in
(\ref{f1}) leads to (\ref{fasyn}).\\
\indent
Let us now discuss how the asymptotic behavior of $g(t)$ determines whether or not finite first-passage times exist.
We first consider a power law behaviour of $g$ at large $t$,
like in the Erd\'elyi-Kober derivative
\begin{equation}
	g(t) \sim t^{\beta}, \hspace{1cm} t \to \infty.
\end{equation}
%corresponding to $g^{-1}(t) \sim t^{1/\beta}$.
We have, from Eq.(\ref{fasyn}), that 
the asymptotic form of the first-passage distribution is
\begin{equation}
	f(t) \sim A\ t^{- (\beta  \alpha + 1)}, \hspace{1cm} t \to \infty.
 \label{f3asyn}
\end{equation}
The condition for the existence of finite MFPT (\ref{condex}) is satisfied for
\begin{equation}
\beta \alpha > 1.
\label{cond1}
\end{equation}
In general, the condition for the existence of finite $k$-th moment of the first-passage time distribution is 
\begin{equation}
\beta \alpha > k .   
\end{equation}
We can then conclude that a finite MFPT for g-fractional diffusion with derivative order $\alpha$ is possible whenever the funtion $g$ diverges at long time faster that $t^{1/\alpha}$. Moreover, finite moments up to the $k$-th are possible if the divergence is faster than $t^{k/\alpha}$.

\noindent
It is worth noting that in the case of anomalous diffusion described by the classical Caputo derivative, i.e. $\beta=1$, the condition (\ref{cond1})
is never satisfied, resulting in a divergent MFPT. The same is true for the case of Hadamard fractional derivative, $g=\ln(t+1)$. Instead, in the case of exponential behavior of $g$ function, $g \sim \exp(\gamma t)$, one has not only a finite MFPT, but finite moments of all orders regardless the value of the derivative order $\alpha$.
We report in Fig.\ref{fig1} the first passage-time distributions for different choices of the $g$ function, highlighting how the asymptotic behavior determines the existence of finite MFPTs in the various cases.

%%%%%%%%%%%%%%%%%%%%%%%%
%  NUOVA 
%%%%%%%%%%%%%%%%%%%%%%%%

\subsection{Fractional diffusion with one absorbing boundary}
The case of g-fractional diffusion in an semi-bounded domain 
$[-a,+\infty$], with one absorbing point at $x=-a$, can be
obtained by taking the limit $b\to \infty$ in the expressions derived in the previous section related to the finite domain case.
The solution of the g-fractional diffusion equation is then obtained from (\ref{ufin}) in the limit $L=a+b \to \infty$.
In such a limit the sums become integrals
\begin{equation}
\frac{\pi}{L} \sum_{n=0}^{\infty} h(\pi n/L) \to 
\int_0^\infty dk \ h(k) ,
\end{equation}
and the solution reads
\begin{equation}
	u(x,t) = 
\frac{2}{\pi} \int_{0}^\infty 
dk \sin[k (x_0+a)]\ \sin[k(x+a)]
\ E_\alpha(-k^2 D g(t)^\alpha) .
\label{uxt2}
\end{equation}
For $g(t)=t$ we recover the expression obtained in \cite{ranga} (Eq.(3.9)).
In the same way we obtain the expression of the first-passage time distribution, taking the limit of (\ref{FPTD})
\begin{equation}
f(t) = \frac{2 D g'(t) g(t)^{\alpha-1}}{\pi}
 \int_{0}^{\infty} dk \ k \ \sin[k(a+x_0)] 
 \ E_{\alpha,\alpha}(-k^2 D g(t)^{\alpha}) .
\label{FPTD2}
\end{equation}
To study the long time behavior of the FPTD we proceed as follows. By changing the integration variable $k \to k g^{\alpha/2}$, using the property (\cite{ml}, p. 86)
\begin{equation}
E_{\alpha,\alpha}(-x) = -\alpha \frac{d}{dx} E_{\alpha}(-x) ,
\end{equation}
and integrating by parts, we rewrite (\ref{FPTD2}) as
\begin{equation}
f(t) = \frac{(a+x_0) \alpha g'(t)}{\pi g(t)^{1+\alpha/2}}
 \int_{0}^{\infty} dk \ 
 \cos\bigg[\frac{k (a+x_0)}{g(t)^{\alpha/2}} \bigg]
 %\cos[k(a+x_0)/g(t)^{\alpha/2}] 
 \ E_{\alpha}(-k^2 D) .
\label{FPTD3}
\end{equation}
Now, introducing the function $H_{\alpha}(x)$, the inverse Laplace transform of $E_{\alpha}(-k)$ for $x>0$
(\cite{ml} p. 92, \cite{Berbe} p. 631)
\begin{equation}
H_{\alpha}(x) = \frac{2}{\pi} \int_0^\infty dk \ \cos(kx)
\ E_{2\alpha}(-k^2) ,
\label{Hfun}
\end{equation}
we can write Eq.(\ref{FPTD3}) as follows
\begin{equation}
f(t) = \frac{(a+x_0) \alpha g'(t)}{2 D^{1/2} g(t)^{1+\alpha/2}}
\ H_{\alpha/2} \Bigg( \frac{a+x_0}{\sqrt{Dg(t)^\alpha}} \Bigg).
\label{FPTD4}
\end{equation}
We can express $H$ as a power series
\begin{equation}
\label{Hserie}
H_{\alpha}(x) = \frac{1}{\pi} \sum_{n=0}^{\infty} c_n(\alpha)
\ x^n ,
\end{equation}
where the coefficients $c_n$ turn out to be \cite{MP2003}
\begin{equation}
c_n(\alpha) = \frac{\pi (-1)^n}{n! \Gamma(1-\alpha-\alpha n)} .
\end{equation}
The FPTD can then be written as
\begin{equation}
f(t) = \frac{(a+x_0) \alpha g'(t)}{2\pi D^{1/2} g(t)^{1+\alpha/2}}
\sum_{n=0}^{\infty} c_n(\alpha/2)
\Bigg( \frac{a+x_0}{\sqrt{Dg(t)^\alpha}} \Bigg)^n .
\label{FPTD5}
\end{equation}
By noting that $b_0(\alpha)= \pi / \Gamma(1-\alpha) \neq 0$ \cite{Berbe}, we have that asymptotically the FPTD behaves as 
\begin{equation}
f(t) \sim g'(t) \ g(t)^{-(1+\alpha/2)} .
\label{f4asyn}
\end{equation}
For $g(t)=t$ the expression (\ref{FPTD5}) and its asymptotic limit $f(t) \sim t^{-1-\alpha/2}$ are in agreement with \cite{ranga} (Eq.s (3.23) and (3.34)).
We note that the asymptotic behavior of FPTD (\ref{f4asyn}) is similar to that obtained for the finite domain case with two absorbing boundaries (\ref{fasyn})
with the substitution $\alpha \to \alpha/2$.
We can then repeat the arguments of the previous section with the rescaled derivative exponent. In particular we have that the MFPT is finite in the case of Erdélyi-Kober derivative $g(t) \sim t^{\beta}$ for $\beta \alpha > 2$ and for the case of exponential $g$ function $g(t) \sim \exp(\gamma t)$. Instead, the MFPT is not defined for $\beta \alpha < 2$ in the Erdélyi-Kober case and in the Hadamard case $g(t)=\ln(t+1)$.
We therefore conclude that, even in the case of semi-infinite domains, for which the MFPT is undefined in Caputo fractional diffusion and also in classical diffusion processes, particular choices of the g-function can lead to finite MPFT.

\subsection{Fractional diffusion in unbounded domains}
We conclude by obtaining the solution $u(x,t)$ of the g-fractional diffusion in an unbounded domain $[-\infty,+\infty]$ as a limit 
$a \to \infty$ of the expression (\ref{uxt2}) derived in the previous section. 
In this limit, using the fact that
$$2 \sin x \sin y = \cos(x-y) - \cos(x+y),$$
Eq.(\ref{uxt2}) becomes
\begin{eqnarray}
u(x,t) &=& \frac{1}{\pi} \int_0^\infty dk \ \cos[k(x-x_0)] \ 
E_\alpha(-k^2 D g(t)^\alpha)  \\
&-& \lim_{a\to\infty} 
\frac{1}{\pi} \int_0^\infty dk \ \cos[k(x+x_0+2a] \ 
E_\alpha(-k^2 D g(t)^\alpha).
\nonumber
\end{eqnarray}
We now show that the second term in the RHS vanishes.
Indeed, we can write the second term as
\begin{equation}
\lim_{a \to \infty} \frac{1}{2 \sqrt{D g(t)^\alpha}}
\ H_{\alpha/2} \Bigg( \frac{x+x_0+2a}{\sqrt{Dg(t)^\alpha}} \Bigg),
\label{lim1}
\end{equation}
where $H_\alpha(x)$ is defined in (\ref{Hfun}).
Now, asymptotically the function $H$ behaves as \cite{MP2003}
\begin{equation}
H_{\alpha/2}(x) \sim x^{-\gamma} \exp{(-\delta x^{\epsilon})},
\end{equation}
where
\begin{eqnarray}
\gamma &=& \frac{1-\alpha}{2-\alpha}, \nonumber\\
\delta &=& (2-\alpha)\ 2^{-\frac{2}{2-\alpha}}
\ \alpha^{\frac{\alpha}{2-\alpha}} , \nonumber\\
\epsilon &=& \frac{2}{2-\alpha}. \nonumber
\end{eqnarray}
The limit (\ref{lim1}) is therefore null and the solution $u(x,t)$ 
reads
\begin{eqnarray}
u(x,t) &=& \frac{1}{\pi} \int_0^\infty dk \ \cos[k(x-x_0)] \ 
E_\alpha(-k^2 D g(t)^\alpha)  \\
&=& 
\frac{1}{2 \sqrt{D g(t)^\alpha}}
\ H_{\alpha/2} \Bigg( \frac{|x-x_0|}{\sqrt{Dg(t)^\alpha}} \Bigg),
\nonumber
\end{eqnarray}
or, using the power series (\ref{Hserie}),
\begin{equation}
u(x,t) = \frac{1}{2 \sqrt{D g(t)^\alpha}} 
\sum_{n=0}^{\infty} \frac{1}{n! \Gamma{(1-\alpha/2 -n\alpha/2)}}
\Bigg(- \frac{|x-x_0|}{\sqrt{Dg(t)^\alpha}} \Bigg)^n ,
\end{equation}
in agreement with Eq.(14) in \cite{kosto}.

%%%%%%%%%%%%%%%%%%%%%%%

%%%%%%%%%%%%%%%%%%%%%%%%
\section{Conclusions}
In this work we have investigated the g-fractional diffusion in bounded 
and semi-bounded
1D domains with absorbing boundaries,
finding explicit solutions of the fractional diffusion equation with derivative of order $\alpha\in(0,1)$ and generic $g$ functions. We focused on first passage time processes,
reporting the exact expression of the first passage time distribution and analyzing the conditions on function $g$ for the existence of finite mean-first passage time and general moments of FPTD. 
We find that finite MFPT is obtained whenever the function $g$ grows faster than $t^{1/\alpha}$ 
(for bounded domains) and
$t^{2/\alpha}$ (for semi-bounded domains)
, and, in general, finite $k$-th moments exist for $g$ functions growing faster than $t^{k/\alpha}$
(bounded) and $t^{2k/\alpha}$ (semi-bounded).

According to the recent paper \cite{kosto}, the function g controls the anomalous diffusion at intermediate
times, with potentially wide application in modeling diffusion processes with variable parameters. With this paper, we have shown the key-role played by 
the choice of the function $g$ in discriminating processes with finite or infinite MFPT.

\section*{Acknowledgments}
L.A. acknowledge financial support from the Italian Ministry of University and Research (MUR) under the PRIN2020 Grant No. 2020PFCXPE.
%L.A. acknowledges financial support from the MIUR PRIN2020 project 2020PFCXPE.
%%%%%%%%%%%%%%%%%%%%%%%%%%%%%%%%%%%%%%%%%%%%%%%%55     

    \end{document}